\documentclass[aps, 10pt]{revtex4}
\usepackage{graphicx,subfigure}
\usepackage{epsfig}
\usepackage{amsmath}
\usepackage{amssymb}
\usepackage{amsthm}
\usepackage{bbold}
\begin{document}
	\title{Metamaterials and Ces\`{a}ro convergence }
	\author{Yuganand Nellambakam and  K. V. S. Shiv Chaitanya }
	\email[]{ chaitanya@hyderabad.bits-pilani.ac.in}
	\affiliation{Department of Physics, BITS Pilani, Hyderabad Campus, Jawahar Nagar, Shamirpet Mandal,
		Hyderabad, India 500 078.}
\begin{abstract}
	In this paper, we show that the linear dielectrics and magnetic materials in matter obey a special kind of mathematical property known as Ces\`{a}ro convergence. Then, we also show that the analytical continuation of the linear permittivity \& permeability to a complex plane in terms of Riemann zeta function. The metamaterials are fabricated materials with a negative refractive index. These materials, in turn, depend on permittivity \& permeability of the linear dielectrics and magnetic materials. Therefore, the Ces\`{a}ro convergence property of the linear dielectrics and magnetic materials may be used to fabricate the metamaterials.
\end{abstract}

\maketitle

\section{Introduction}
The discovery of artificial media, that is, metamaterials in the last decade, has given rise to different kinds of phenomena such as perfect lens \cite{len}, Metamaterial Antennas \cite{ma}, Clocking Devices \cite{clo}, Acoustic Metamaterials \cite{am}, Seismic Metamaterials \cite{sm} which are not exhibited by existing natural materials. 
The materials which are characterized by negative refractive index metamaterials. The seminal paper on negative refractive index materials for the non-dissipative mediums was proposed Veselago \cite{vso},   where both permittivity and permeability are simultaneously negative. Further, he has shown electromagnetic wave propagation in negative refractive index materials exhibits a unique property such as phase velocity is antiparallel to the direction of energy flow, the reversal of the Doppler effect and Cerenkov radiation.  In an isotropic dielectric–magnetic medium with dissipation, a general condition for phase velocity directed oppositely to the power flow was derived by Lakhtakia \cite{lak, lak1, lak2}. In the same paper, it has been shown that the real parts of both the permittivity and the permeability need not be both negative. Valanju et al. \cite{val} have shown that the group fronts refract positively even when phase fronts refract negatively for a negative index material. In the references  \cite{kk,kk1,kk2}, the relationship between the Kramers-Kronig relations and negative index of refraction is investigated.  From the above discussion, it is clear that for material in a dispersive medium, the refractive is complex.  Therefore, the refractive index being complex constraints permittivity and permeability also to be complex. The refractive index is defined as $n_r=\sqrt{\epsilon_r\mu_r}$, where $n_r$ is refractive index, $\epsilon=\epsilon_{0}\epsilon_r$ is permittivity here $\epsilon_0=8.85\times 10^{-12}\frac{C^2}{N M^2}$ is permittivity of free space, $\epsilon_r$ is relative permittivity and $\mu=\mu_{0}\mu_r$ here $\mu_0=4\pi \times 10^{-7}N/A^2$ permeability of free space  and $\mu_r$ is relative permeability.  For the sake of brevity, we take $\epsilon_0=\mu_0=1$ throughout this paper. 

The electric field inside a sphere of homogeneous linear dielectric material is placed in an otherwise uniform electric field $E_0$ can be calculated by two approaches, first as a boundary value problem, and the second method is by using successive approximations \cite{grif}. But, we encounter an issue when using the method of successive approximations is that the result is not valid for all values of permittivity. A similar situation arises when a  linear spherical magnetic material placed in a uniform magnetic field $\textbf{B}_{0}$, which can also be calculated by two approaches, first as a boundary value problem and the second method is by using successive approximations \cite{grif}. Therefore, in this paper, we address this problem by showing that linear dielectrics and magnetic materials in matter obey a special kind of mathematical property is known as Ces\`{a}ro convergence. This Ces\`{a}ro convergence enables to extend the results to all values of permittivity by analytical continuation to a complex plane in terms of Riemann zeta function. The metamaterials are fabricated materials with a negative refractive index. These materials, in turn, depend on permittivity \& permeability of the linear dielectrics and magnetic materials. Therefore, the Ces\`{a}ro convergence property of the linear dielectrics and magnetic materials may be used to fabricate the metamaterials

The paper is organized as follows: in section II, we give a brief introduction to the origin of the problem that is the electric field inside a sphere of homogeneous linear dielectric and spherical magnetic materials placed in a uniform magnetic field and two approaches. In section III, we discuss our solution in terms of Ces\`{a}ro convergence. In section IV, we show that the Ces\`{a}ro convergence leads to the analytical continuation to a complex plane in terms of Riemann zeta function. Finally, we conclude the paper in section V.

\section{Linear permittivity \& permeability} 
The electric field inside a sphere of homogeneous linear dielectric material is placed in
an otherwise uniform electric field $E_0$ is given by \cite{grif}
\begin{equation}\label{e1}
E=\frac{3}{\epsilon_r + 2} E_0
\end{equation}
where $E_0$ is the inside field. One can also arrive at the same result by the following method of successive approximations that is by considering the initial field inside the sphere is  $\textbf{E}_0$, which give rise to the polarization $\textbf{P}_0 = \epsilon_{0}\chi_{e}\textbf{E}_0$. This polarization $\textbf{P}_0$ generates a field of its own, say, $\textbf{E}_1=-\frac{1}{3\epsilon_0}P_0=-\frac{\chi_e}{3}E_0$, which in turn modifies the polarization by an amount $\textbf{P}_1=\epsilon_0\chi_eE_1=-\frac{\epsilon_0\chi_e^2}{3}E_0$, which further generates the field $\textbf{E}_2=-\frac{1}{3\epsilon_0}P_1=\frac{\chi_e^2}{9}E_0$, and so on. Therefore, the resulting field is given by \cite{grif}
\begin{equation}\label{e2}
E=E_0 + E_1 + E_2 + · · ·=\sum_{0}^{\infty}\left(-\frac{\chi_e}{3}\right)^nE_0
\end{equation} 
It is clear that the equation (\ref{e2}) is a geometric series and summed explicitly:
\begin{equation}\label{e3}
E=\frac{1}{1+\frac{\chi_e}{ 3}}E_0=\frac{3}{\epsilon_r + 2} E_0
\end{equation}
where $\epsilon_r=1+\chi_e$ which agrees with equation (\ref{e1}). Readers should note that geometric series of the form $\sum_{0}^{\infty}x^n=\frac{1}{1-x}$ is valid only for values of $-1<x<1$, here $-1$ and $1$  are also excluded. Therefore, in the case of equation (\ref{e3}) requires that $\chi_e < 3$ else the infinite series
diverges. But from equation (\ref{e1}) the result is subject to no such restriction hence equations (\ref{e1}) and (\ref{e2}) are inconsistent. For example $\chi_e=3$
the series\begin{equation}\label{e24}
E=E_0 + E_1 + E_2 + · · ·=\sum_{0}^{\infty}\left(-\frac{\chi_e}{3}\right)^nE_0
\end{equation} reduces to
\begin{equation}\label{e21a}
E=(1-1+1-1+1.......)E_0
\end{equation}
In literature, the series $1-1+1-1+1.......$  is known as Grandi's series and is divergent.

Similarly, the Magnetic field of a  linear spherical magnetic material placed in an uniform magnetic field $\textbf{B}_{0}$ is given by \cite{grif}
\begin{equation}\label{b0}
\textbf{B}=\mu\textbf{H}=\frac{3\textbf{B}_{0}}{(2\mu_{0}+\mu)}=\left(\frac{1+\chi_m}{1+\chi_m/3}\right)\textbf{B}_0.
\end{equation}
One can also arrive at the same result   by the following method of successive approximations that is we consider the initial field inside the sphere is  $\textbf{B}_0$ magnetizes the sphere: $M_0$=$\chi_m$$\textbf{H}_0$=$\frac{\chi_m}{\mu(1+\chi_m)}$$\textbf{B}_0$. This magnetization sets up a field within the sphere $
\textbf{B}_1=\frac{2}{3}\mu_0\textbf{M}_0=\frac{2}{3}\frac{\chi_m}{1+\chi_m}\textbf{B}_0=\frac{2}{3}\kappa\textbf{B}_0$, where $\kappa=\frac{\chi_m}{1+\chi_m}$,
which in turn modifies the magnetizes of sphere by an additional amount $\textbf{M}_1$=$\frac{\kappa}{\mu_0}$$\textbf{B}_1$. This sets up an additional field in the sphere $ \textbf{B}_2=\frac{2}{3}\mu_0\textbf{M}_1=\frac{2}{3}\kappa\textbf{B}_1=\left(\frac{2\kappa}{3}\right)^2\textbf{B}_0,$
and so on. Therefore the resulting field is given by \cite{grif}
\begin{eqnarray}
\textbf{B}&=&\textbf{B}_0+\textbf{B}_1+\textbf{B}_2+...\\&=&\textbf{B}_0+(2\kappa /3)\textbf{B}_0+(2\kappa /3)^2\textbf{B}_0+....\label{b1}\\
&=&[1+(2\kappa /3)+(2\kappa /3)^2+....]\textbf{B}_0\label{c2}\\
&=&\frac{\textbf{B}_0}{(1-2\kappa/3)}=\frac{3\textbf{B}_0}{3-2\chi_m/(1+\chi_m)}\\&=&\frac{(3+3\chi_m)\textbf{B}_0}{3+3\chi_m-2\chi_m}
=\left(\frac{1+\chi_m}{1+\chi_m/3}\right)\textbf{B}_0.\label{b2}
\end{eqnarray}

For the value of $\chi_m=-\frac{3}{5}$ gives $\kappa=-\frac{3}{2}$ the equation (\ref{e1}) reduces to
\begin{equation}\label{e21b}
\textbf{B}=(1-1+1-1+1.......)\textbf{B}_0
\end{equation}
Again we arrive at Grandi's series. It should be noted that equations (\ref{b0}) and (\ref{b2}) are inconsistent.

\section{Ces\`{a}ro Convergence }
It should be noted that  equations (\ref{e1}) and (\ref{e3})and equations (\ref{b0}) and (\ref{b2}) are inconsistent as equation (\ref{e1}) and equation (\ref{b0}) are valid for the all values of $\chi_{e}$ and $\chi_m$ respectively. Hence, the equations (\ref{e3}) and (\ref{b2}) also should be valid for all values of $\chi_{e}$ and $\chi_m$ respectively. One of the cases, where equations (\ref{e3}) and (\ref{b2}) diverges for  $\chi_{e}=-3$ and $\chi_m=-\frac{3}{5}$ and  the series generated are given by equations   (\ref{e21a}) and (\ref{e21b})  respectively are known as Grandi's series  in literature. The Grandi's series Q is in general is a divergent series is given by
\begin{equation}\label{cs1}
Q=1-1+1-1+1.......=\sum_{n=0}^{+\infty}Q_j=\sum_{n=0}^{+\infty}(-1)^n
\end{equation}
But, this series converges to a value of $\frac{1}{2}$ for a special kind of convergence known as the Ces\`{a}ro convergence. In a geometric series the sequence of partial sums converges to a real number, for a series which obeys Ces\`{a}ro sum the average of partial sums converges to a real number.  The  Ces\`aro sum is defined as
\begin{equation}\label{def}
\sigma_n = \sum_{j=0}^{n-1}(1-\frac{j}{n})a_j =  \frac{s_0+s_1+.. s_{n-1}}{n}=\frac{1}{n}\sum_{j=0}^{n-1}s_j
\end{equation}
A series $\sum_{j=0}^{n} a_j $ is called Ces\`aro summable satisfies the following theorem:

\textbf{Theorem 1} Suppose that $\sum_{j=0}^{n} a_j $ is a convergent series with sum, say L. Then $\sum_{j=0}^{n} a_j $ is Ces\`aro summable to L. 
\begin{equation}\label{def1}
\lim_{n\rightarrow\infty}s_n = L \in \mathbb{R}\;\;\;\;\Rightarrow\;\;\;\lim_{n\rightarrow\infty}\sigma_{n} = L \in \mathbb{R}.
\end{equation}
The proof is given in \cite{cso}. Following are the properties of  Ces\`aro sums:
If $\sum_n a_n$ = A and $\sum_n b_n$ = B are convergent series, then \\
i. Sum-Difference Rule:  $\sum_n(a_n\pm b_n)$ = $\sum_n a_n \pm \sum _n b_n$ = A $\pm$ B \\
ii. Constant Multiple Rule : $\sum_n c\; a_n$ = c $\sum_n a_n$ = cA for any real number c.\\
iii. The product of $AB=\sum_n a_n\sum_n b_n$ is also as Ces\`aro sums. 

The sequence in equation (\ref{cs1}) has two possibilities; one case is sequence ends with an even number of terms, and the other argument is series ends with an odd number of terms.  For the even of terms, the partial sums add to $s_{2n} = 0$, and for the odd number of terms, the partial sums add to $s_{2n+1} = 1$, then the average of the even and odd is $1/2$. Then, by followings the  definition (\ref{def})
\begin{eqnarray}
\sigma_{2n+1} &=& \frac{1}{2n+1}(1+0+1+0+...+1)=\frac{n+1}{2n+1} \label{odd}\\
\sigma_{2n} &=& \frac{1}{2n}(1+0+1+0+...+0) =\frac{1}{2}\label{even}
\end{eqnarray}
The readers should not that the terms of sequence  in the equations (\ref{odd}) and (\ref{even})  are generated from equation $\ref{cs1}$ as
\begin{eqnarray}
Q_0=1,;\;\;\;\ Q_0+Q_1=0,;\;\;\;\ Q_0+Q_1+Q_2=1, ;\;\;\;\ Q_0+Q_1+Q_2+Q_3=0,......
\end{eqnarray}
Then by applying the theorem1 we get 
\begin{equation}
\lim_{n\rightarrow\infty}\; \sigma_{2n} = \lim_{n\rightarrow\infty}\sigma_{2n+1} = \frac{1}{2}.
\end{equation}

The refractive index $n$ is defined as $n_r=\sqrt{\epsilon_r\mu_r}$, where $\epsilon_r$ is relative permittivity and $\mu_r$ relative permeability. Since the  $\epsilon_r$ is relative permittivity and $\mu_r$ relative permeability both satisfy the sequence given in equation (\ref{cs1}) we get refractive index to be 
\begin{equation}\label{ref}
n_r^2=\mu\nu=(1-1+1-1+....)^2
\end{equation} The sequence (\ref{ref}) satisfies the following identity
\begin{eqnarray}
n_r=(1-1+1-1+....)^2 &=& (1-1+1-1+...) x (1-1+1-1+...)\nonumber\\
&=& 1-2+3-4+5+.....\sum_{j=0}^\infty(-1)^jj.\label{alt}
\end{eqnarray}
By applying the third property of Ces\`aro sums that is the product of $AB=\sum_n a_n\sum_n b_n$ is also as Ces\`aro sums one gets
\begin{equation}\label{rty}
n_r=(1-1+1-1+....)^2= (1/2)^2 = 1/4.
\end{equation}
Thus, we have shown that the equations (\ref{e1}) and (\ref{e3}) and the equations (\ref{b0}) and (\ref{b2}) are valid for the all values of $\chi_{e}$ linear  permittivity and $\chi_m$   linear permeability as the equations (\ref{e3}) and (\ref{b2}) obeys Ces\`{a}ro convergence, in the processes linear  permittivity \& permeability becomes complex and we will discuss in next section.

\section{Analytical continuation to Riemann Zeta($\zeta$) function}
As pointed out in the previous section linear  permittivity \& permeability in  equations (\ref{e1}) and (\ref{b0}) will become complex  as , as Ces\`{a}ro sum is defined for complex number \cite{sas}. For example consider the geometric series in equation (\ref{e2}) by treating $\chi_e$ to be complex then the value of $\chi_e=-3$ is falls inside the domain of convergence \cite{sas}.  This process is known as the analytical continuation. One of the well-known examples of analytical continuation is the Riemann Zeta ($\zeta$) function and is given by
\begin{equation}
\zeta(s)=\sum_{k=1}^{+\infty}\frac{1}{k^s}=1+\frac{1}{2^s}+\frac{1}{3^s}+\frac{1}{4^s}+\frac{1}{5^s}+....
\end{equation}
The Riemann Zeta is defined for the value $s>1$, and for all other values below less than or equal to one on the real line, it diverges. Riemann has extended the domain of convergence of $\zeta$ function to the entire real line by allowing $s=\sigma+i\omega$ to be complex  and relating it to the Dirichlet eta function
\begin{equation}\label{dr}
\eta(s)=\sum_{k=1}^{+\infty}\frac{(-1)^k}{k^s}=1-\frac{1}{2^s}+\frac{1}{3^s}-\frac{1}{4^s}+\frac{1}{5^s}+......
\end{equation}
and this series converges for the value of $s>1$ and diverges for the value of $s\leq 1$. By allowing $s$ to be complex  the domain of Dirichlet eta function (\ref{dr}) is extended to entire complex plane where the divergent series obeys the Ces\`{a}ro convergence. For example when $s=0$ we recover Grandi's series (\ref{cs1}) and $s=1$ we recover the series in equation (\ref{alt}). The  analytic continuation of $\zeta(s)$ in terms of  Dirichlet Eta Function $\eta(s)$ is given by
\begin{eqnarray}
\eta(s) &=& 1-\frac{1}{2^s}+\frac{1}{3^s}-\frac{1}{4^s}+\frac{1}{5^s}+....\\&=&\left(1+\frac{1}{2^s}+\frac{1}{3^s}+\frac{1}{4^s}+\frac{1}{5^s}+..\right)
-\left(\frac{2}{2^s}+\frac{2}{4^s}+\frac{2}{6^s}+\frac{2}{8^s}+..\right)\nonumber\\
&=& \zeta(s)-\frac{1}{2^{s-1}}\zeta(s)= (1-2^{1-s})\zeta(s)
\end{eqnarray}
For the value of $s=0$ we recover the (\ref{e21a}) and (\ref{e21b})  respectively whose Riemann Zeta function values are $\zeta(0)=-1/2$ and refractive index in equation (\ref{rty}) to be $\zeta(0)^2=1/4$.

\section{Conclusion}
In this paper,  we have shown that the linear dielectrics and magnetic materials in matter obey a special kind of mathematical property known as Ces\`{a}ro convergence. Then, we have also shown that the analytical continuation of the linear permittivity \& permeability to complex plane in terms of Riemann zeta function. The metamaterials are fabricated materials with a negative refractive index. These materials, in turn, depend on permittivity \& permeability of the linear dielectrics and magnetic materials. Therefore, the Ces\`{a}ro convergence property of the linear dielectrics and magnetic materials may be used to fabricate the metamaterials.
\section*{Acknowledgments}
KVSSC acknowledges the Department of Science and Technology, Govt of India (fast-track scheme (D. O. No: MTR/2018/001046)), for financial support.

\end{document}